\documentclass[twocolumn]{aastex61}
\usepackage{amsmath}

\submitjournal{ApJ}
\accepted{July 17, 2019}

\shorttitle{Low Metallicity in 3C368}
\shortauthors{Lamarche et al.}

\begin{document}

\title{CO and Fine-Structure Lines Reveal Low Metallicity in a Stellar-Mass-Rich Galaxy at \lowercase{z} $\sim$ 1?}

\correspondingauthor{Cody Lamarche}
\email{cjl272@cornell.edu}

\author{C. Lamarche}
\affil{Department of Astronomy, Cornell University, Ithaca, NY 14853.}

\author{G. J. Stacey}
\affil{Department of Astronomy, Cornell University, Ithaca, NY 14853.}

\author{A.Vishwas}
\affil{Department of Astronomy, Cornell University, Ithaca, NY 14853.}

\author{D. Brisbin}
\affil{N$\acute{\rm{u}}$cleo de Astronom$\acute{\rm{i}}$a, Facultad de Ingenier$\acute{\rm{i}}$a y Ciencias, Universidad Diego Portales, Av. Ej$\acute{\rm{e}}$rcito 441, 8370191 Santiago, Chile.}

\author{C. Ferkinhoff}
\affil{Department of Physics, Winona State University, Winona, MN, 55987.}

\author{T. Nikola}
\affil{Cornell Center for Astrophysics and Planetary Science, Cornell University, Ithaca, NY 14853.}

\author{S. J. U. Higdon}
\affil{Department of Physics, Georgia Southern University, Statesboro, GA 30460.}

\author{J. Higdon}
\affil{Department of Physics, Georgia Southern University, Statesboro, GA 30460.}

\begin{abstract}

We present detections of the CO(4-3) and [C\,{\sc i}]\,609\,$\micron$ spectral lines, as well as the dust continuum at 480.5\,GHz (rest-frame), in 3C\,368, a Fanaroff-Riley class II (FR-II) galaxy at redshift (z) 1.131. 3C\,368 has a large stellar mass, $\sim$\,3.6\,$\times$\,10$^{11}$\,M$_\odot$, and is undergoing an episode of vigorous star formation, at a rate of $\sim$\,350\,$M_{\odot}$\,yr$^{-1}$, and active galactic nucleus (AGN) activity, with radio-emitting lobes extended over $\sim$\,73 kpc. Our observations allow us to inventory the molecular-gas reservoirs in 3C 368 by applying three independent methods: (1) using the CO(4-3)-line luminosity, excitation state of the gas, and an $\alpha_{CO}$ conversion factor, (2) scaling from the [C\,{\sc i}]-line luminosity, and (3) adopting a gas-to-dust conversion factor. We also present gas-phase metallicity estimates in this source, both using far-infrared (FIR) fine-structure lines together with radio free-free continuum emission and independently employing the optical [O\,{\sc iii}]\,5007\,\AA\,and [O\,{\sc ii}]\,3727\,\AA\,lines (R$_{23}$ method). Both methods agree on a sub-solar gas-phase metallicity of $\sim$\,0.3\,Z$_\odot$. Intriguingly, comparing the molecular-gas mass estimated using this sub-solar metallicity, $M_{gas}$\,$\sim$\,6.4\,$\times$\,10$^{10}$\,$M_{\odot}$, to dust-mass estimates from multi-component spectral energy distribution (SED) modeling, $M_{dust}$\,$\sim$\,1.4\,$\times$\,10$^{8}$\,$M_{\odot}$, yields a gas-to-dust ratio within $\sim$\,15\% of the accepted value for a metallicity of 0.3\,Z$_\odot$. The derived gas mass puts 3C\,368 on par with other galaxies at z\,$\sim$\,1 in terms of specific star-formation rate and gas fraction. However, it does not explain how a galaxy can amass such a large stellar population while maintaining such a low gas-phase metallicity. Perhaps 3C\,368 has recently undergone a merger, accreting pristine molecular gas from an external source.

\end{abstract}

\keywords{galaxies: evolution – galaxies: high-redshift – galaxies: ISM – galaxies: star formation – ISM: photon-dominated region (PDR) – ISM: H {\sc ii} regions}

\section{Introduction}

Recent studies have made great strides in inventorying molecular gas, the fuel for star formation, in galaxies during the peak epoch of cosmic star formation, redshift (z) $\sim$ 1-3, in an effort to determine the efficiency with which these galaxies convert their molecular gas reservoirs into stars \citep[e.g.,][]{Tacconi2010, Harris2012, Sharon2016, Harrington2018, Pavesi2018}. These surveys are complicated by the fact that hydrogen, whose lowest rotational transition takes $\sim$ 500 K to excite, is difficult to observe in cold molecular clouds, so that the easily-excited low-J rotational levels of carbon-monoxide (CO) are generally used as a proxy for molecular gas content. Under some interstellar medium (ISM) conditions however, appreciable amounts of molecular hydrogen are not accompanied by CO. These so-called ``CO-dark" molecular clouds are troublesome in that they cause galaxies containing them to appear very gas poor when employing CO-derived gas-mass estimates. Several possible explanations for these CO-dark molecular clouds exist, including low-metallicity gas present within the ISM of galaxies \citep[e.g.,][]{Maloney1988, Stacey1991} and highly-fractionated molecular clouds \citep[e.g.,][]{Storzer1997}. Here we investigate one such CO-dark galaxy, 3C 368 \citep[e.g.,][]{Evans1996, Lamarche2017}, which has an appreciable star-formation rate and seemingly very little molecular gas.

3C 368 was discovered as part of the Third Cambridge Radio Catalog \citep{Edge1959} and has come to be known as one of the archetypal Fanaroff-Riley class II (FR-II) galaxies. It is an interesting source, situated within the peak epoch of cosmic star-formation, forming stars at a rate of $\sim$ 350 $M_{\odot}$ yr$^{-1}$ \citep{Podigachoski2015}, and containing an active galactic nucleus (AGN) that has launched radio-emitting lobes which span $\sim$ 73 kpc \citep{Best1998}.

In \cite{Lamarche2017} we used both mid- and far-infrared (FIR) fine-structure line observations of 3C 368 to infer the age of the young stellar component present within the source, finding an age of $\sim$ 6.5 Myrs, consistent with that determined using a multi-component model for the observed UV-FIR spectral energy distribution (SED) \citep{Drouart2016}. We also presented a non-detection of the CO(2-1) line, constraining the CO luminosity to a level twelve times lower than expected in star-forming galaxies, based on standard ratios with the [C\,{\sc ii}] 158 $\micron$ line \citep[e.g.,][]{Stacey1991}, which we attributed to either a low-metallicity or highly-fractionated interstellar medium (ISM). In this follow-up paper, we present observations of the CO(4-3) and [C\,{\sc i}](1-0) 609 $\micron$ spectral lines, which emanate from molecular clouds and photo-dissociation regions (PDRs), respectively, as well as the dust continuum emission at 480.5\,GHz (rest-frame), in 3C 368, and show how these measurements help us to disentangle these two competing scenarios.

We assume a flat $\Lambda$CDM cosmology, with $\Omega_M$ = 0.27, $\Omega_\Lambda$ = 0.73, and H$_0$ = 71\,km\,s$^{-1}$\,Mpc$^{-1}$, throughout this paper \citep{Spergel2003}.

\section{Observations and Data Reduction}

\subsection{ALMA}

The CO(4-3) and [C\,{\sc i}] 609 $\micron$ lines were observed simultaneously in 3C 368, using the Atacama Large Millimeter/submillimeter Array (ALMA)\footnote{The National Radio Astronomy Observatory is a facility of the National Science Foundation operated under cooperative agreement by Associated Universities, Inc.} band-6 receivers. The observations were split into two separate execution blocks: the first conducted on June 3, 2016, with a precipitable water vapor (PWV) measurement of 1.24 mm and baselines up to 772 m, and the second conducted on August 20, 2016, with a PWV of 0.31 mm and baselines up to 1462 m. The total on-source integration time for these two observations was $\sim$ 1 hour and 40 minutes.

For both execution blocks, J1751+0939 (PKS 1749 +096) was used as the bandpass, flux, and phase calibrator. The data from the two execution blocks were reduced using the Common Astronomy Software Application (CASA)\footnote{https://casa.nrao.edu/} ALMA pipeline, version 38366 (C4 --- R2B), run in CASA version 4.7.0, and subsequently combined, imaged, and cleaned, using CASA version 4.7.2.

Imaging the CO(4-3) measurement set at a spectral resolution of 60 km s$^{-1}$, and employing a robust parameter of 0.5, we obtain an RMS sensitivity of 120 $\mu$Jy beam$^{-1}$ in each spectral channel, with a synthesized beam of size 0$\farcs$41 $\times$ 0$\farcs$31.

The [C\,{\sc i}] 609 $\micron$ measurement set was similarly imaged at 60 km s$^{-1}$ spectral resolution, employing a robust parameter value of 0.5. Here we obtain an RMS sensitivity of 130 $\mu$Jy beam$^{-1}$ in each spectral channel, with a synthesized beam of size 0$\farcs$39 $\times$ 0$\farcs$29.

A continuum image was created by combining the two line-free spectral windows present in the measurement set, for a total continuum bandwidth of $\sim$\,4\,GHz, centered at 225.5\,GHz (observed-frame). Imaging with a robust parameter value of 0.5, as was done for the line data, we obtain an RMS noise of 13 $\mu$Jy beam$^{-1}$, with a beam of size 0$\farcs$38 $\times$ 0$\farcs$29.

\section{Results and Discussion}

\subsection{Line and Continuum Fluxes}

Both the CO(4-3) and [C\,{\sc i}] 609 $\micron$ lines are well detected in our ALMA observations, emanating from the core component of 3C 368. A continuum-subtracted moment-zero map was created for each of the two spectral lines (see Figure 1), with the corresponding line fluxes determined by fitting a two-dimensional Gaussian to the emitting region of each moment-zero map. We obtain a CO(4-3) line flux of 1.08 $\pm$ 0.14 Jy km s$^{-1}$, or equivalently (7.8 $\pm$ 1.0) x 10$^{-21}$ W m$^{-2}$, and a [C\,{\sc i}] 609 $\micron$ line flux of 0.85 $\pm$ 0.09 Jy km s$^{-1}$, or equivalently (6.6 $\pm$ 0.7) x 10$^{-21}$ W m$^{-2}$ (see Table 1). These fluxes were verified by numerically integrating the spectra extracted from an aperture containing the core of 3C 368 (see Figure 3) and were found to be consistent within the quoted uncertainties.

The continuum in 3C 368 is resolved into three components in the 225.5\,GHz (observed-frame) map: the northern lobe at 226 $\pm$ 59 $\mu$Jy, the southern lobe at 128 $\pm$ 28 $\mu$Jy, and the core at 96 $\pm$ 21 $\mu$Jy (see Figure 2 and Table 1), where these flux densities were determined by fitting a two-dimensional Gaussian to each component in the continuum image. We note that continuum flux extended around the core component of 3C 368 at the same spatial scales as the [C\,{\sc i}] and CO emission could be resolved out at the resolution and sensitivity of our observations, making the quoted flux density from the core a lower limit on the true value.

\begin{figure*}
\gridline{\fig{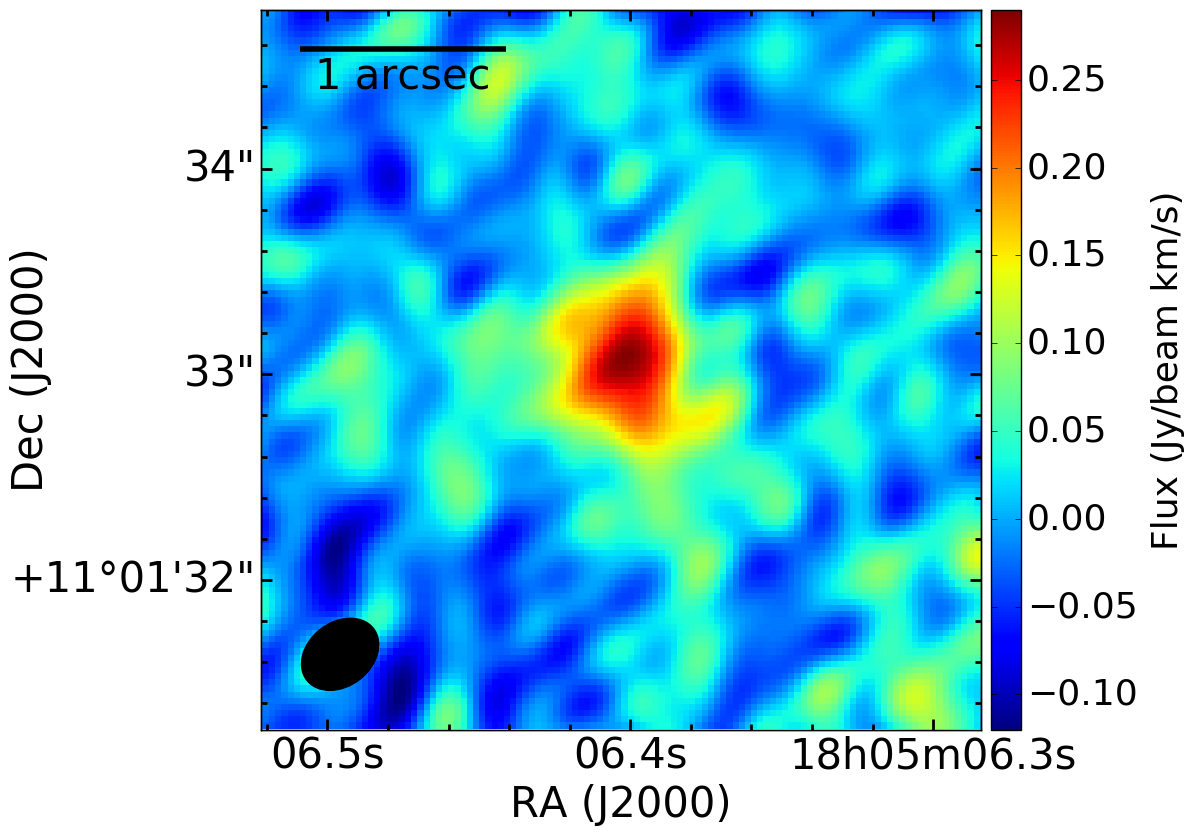}{0.5\textwidth}{(a)}
          \fig{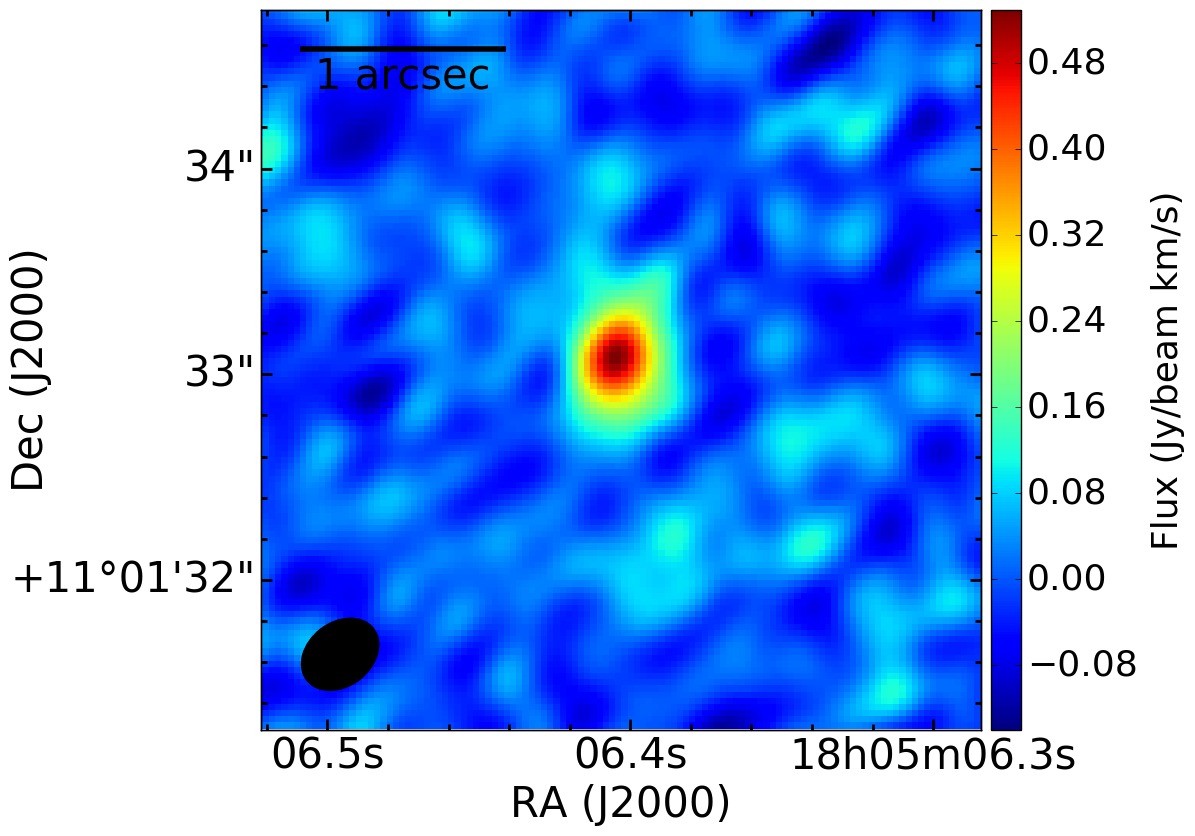}{0.5\textwidth}{(b)}
          }
\caption{(a) ALMA CO(4-3) moment-zero map showing the core component of 3C 368 at 0$\farcs$4 resolution. The CO-emitting source has a size of 0$\farcs$77 $\times$ 0$\farcs$46, deconvolved from the beam. (b) ALMA [C\,{\sc i}] 609 $\micron$ moment-zero map showing the same region of 3C 368 at 0$\farcs$4 resolution. The deconvolved [C\,{\sc i}]-emitting source size is 0$\farcs$43 $\times$ 0$\farcs$23.}
\end{figure*}

\begin{figure}
\plotone{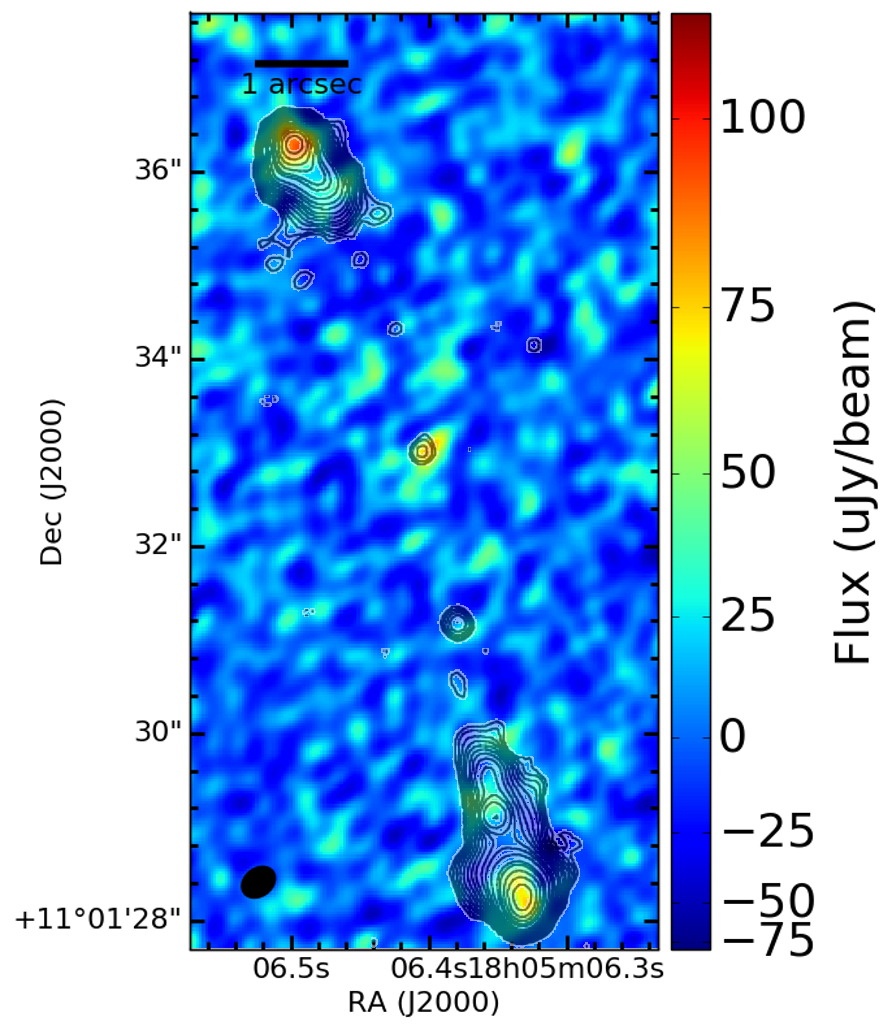}
\caption{ALMA 225.5\,GHz (observed-frame) continuum colormap of 3C 368 at 0$\farcs$4 resolution. The source is resolved into three components: a core, a northern lobe, and a southern lobe, co-spatial with those seen in 8.21 GHz (observed-frame) radio-continuum emission \citep[contour map;][]{Best1998}. The core component of 3C 368 is unresolved at this resolution and sensitivity.}
\end{figure}

\begin{figure}
\plotone{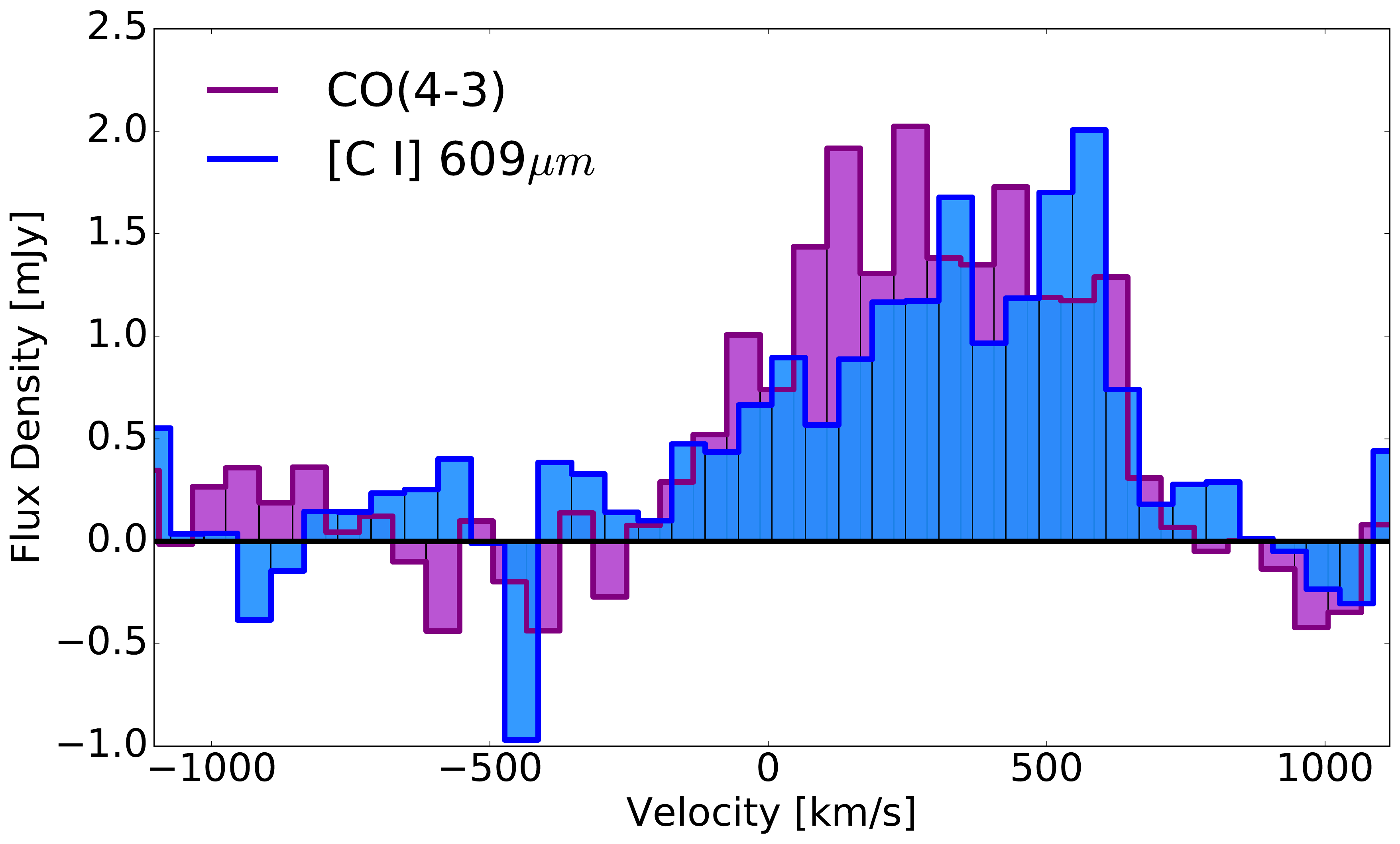}
\caption{Continuum-subtracted ALMA spectra of both the CO(4-3) and [C\,{\sc i}] 609 $\micron$ lines extracted from the core component of 3C 368. v = 0 corresponds to z = 1.131.}
\end{figure}

\begin{deluxetable*}{ccccc}
\tablecaption{Spectral-line and continuum observations of 3C 368}
\tablecolumns{5}
\tablenum{1}
\tablehead{
\colhead{Observation} & \colhead{Core} & \colhead{N. Lobe} & \colhead{S. Lobe} & \colhead{Unit}\\
}
\startdata
CO(4-3) & 7.8\,$\pm$\,1.0 & - & - & F$_{obs}$\,($10^{-21}$\,W\,m$^{-2}$) \\
'' & 4.6\,$\pm$\,0.6 & - & - & L'\,(10$^{9}$\,K\,km\,s$^{-1}$\,pc$^2$) \\\relax
[C\,{\sc i}](1-0) & 6.6\,$\pm$\,0.7 & - & - & F$_{obs}$\,($10^{-21}$\,W\,m$^{-2}$) \\
'' & 3.2\,$\pm$\,0.3 & - & - & L'\,(10$^{9}$\,K\,km\,s$^{-1}$\,pc$^2$) \\
Continuum ($\nu_{obs}$ = 225.5\,GHz) & 96$\pm$21 & 226$\pm$59 & 128$\pm$28 & $S_{\nu, obs}$\,($\mu$Jy) \\
\enddata
\tablecomments{These flux and flux-density values were determined by fitting two-dimensional Gaussian profiles to the respective maps for each of the source components within 3C 368 (see the text).}
\end{deluxetable*}

\subsection{Continuum Science}

\subsubsection{Spectral Energy Distribution}
Using the ALMA continuum observations centered at 480.5\,GHz (rest-frame) presented here, together with our previous ALMA continuum measurements at 205 $\micron$ \citep[rest-frame;][]{Lamarche2017}, we can calculate the slope of the Rayleigh-Jeans power-law for the thermal dust continuum of 3C 368. This is particularly interesting in this case, since the core component of 3C 368, the star-forming galaxy, makes up the smallest contribution to the continuum flux at rest-frame 480.5\,GHz, the flux being dominated instead by the radio lobes. Hence, unresolved continuum measurements at mm/sub-mm wavelengths may have overestimated the flux from the core of 3C 368. We obtain a Rayleigh-Jeans power-law index, S$_\nu$ $\sim$ $\nu^\alpha$, of 3.7 $\pm$ 0.4, or equivalently a dust-emissivity $\beta$ value, S$_\nu$ $\sim$ $\nu^{2 + \beta}$, of 1.7 $\pm$ 0.4. This is consistent with the dust-emissivity $\beta$ value assumed in \cite{Podigachoski2015}, who used that value to derive an FIR luminosity of 2.0 $\times$ 10$^{12}$ $L_{\odot}$, which we used in our previous work and will adopt throughout this paper.

We can also decompose the radio spectral energy distribution (SED) from the core of 3C 368 into thermal and nonthermal components, using our previous ALMA 230 GHz (rest-frame) continuum observations \citep{Lamarche2017}, together with radio continuum observations at 8.21 and 4.71 GHz (observed-frame) \citep{Best1998}, using an equation of the following form \citep[e.g.,][]{Condon1992, Klein2018}:

\begin{equation}
S_{total, r} = S_{th, 0, r} \bigg( \frac{\nu}{\nu_{0,r}} \bigg) ^{-0.1} + S_{nth, 0, r} \bigg( \frac{\nu}{\nu_{0,r}} \bigg) ^{-\alpha_{nth}}
\end{equation}

where $S_{th,0,r}$ and $S_{nth,0,r}$ are the (rest-frame) contributions to the total radio flux from the thermal and nonthermal components, respectively, at $\nu_{0,r}$ (rest-frame), and $\alpha_{nth}$ is the nonthermal power-law index. Adopting a $\nu_{0,r}$ value of 230 GHz and holding $\alpha_{nth}$ fixed at 0.7 \citep[e.g.,][]{ShuBook}, we fit for $S_{th,0,r}$ and $S_{nth,0,r}$. We find best-fit values of $S_{th,0,r}$ = 45 $\pm$ 18 $\mu$Jy and $S_{nth,0,r}$ = 4 $\pm$ 3 $\mu$Jy (see Figure 4). 

We do not consider any contribution to this part of the SED from thermal dust emission, which, continuing the Rayleigh-Jeans power-law found above, would contribute only $\sim$ 5\% of the flux measured at 230\,GHz (rest-frame). Indeed, the continuum flux from the core of 3C 368 is observed to be brighter at rest-frame 230 GHz than at rest-frame 480.5\,GHz, indicating that the dominant emission mechanism at 230\,GHz is not thermal dust -- an inference made possible by the continuum observations presented here.

Beyond the core component of 3C 368, we note that the radio lobes have considerable emission even in the high-frequency observations presented here (rest-frame 480.5\,GHz). Between observed-frame 8.21 and 4.71 GHz, \cite{Best1998} find a spectral index, S$_\nu$ $\sim$ $\nu^\alpha$, of $\sim$ -1.5 for both the northern and southern radio lobes. While the flux densities at rest-frame 480.5\,GHz fall below the extrapolated power-law values, by factors of $\sim$ 1.4 and 2.3 for the northern and southern lobes, respectively, the fact that such energetic electrons are present within the radio-emitting lobes is noteworthy. While detailed modeling is beyond the scope of this paper, the observations presented here suggest that the radio-emitting lobes of 3C 368 are quite young \citep[e.g.,][]{Murgia1999}.

\begin{figure}
\plotone{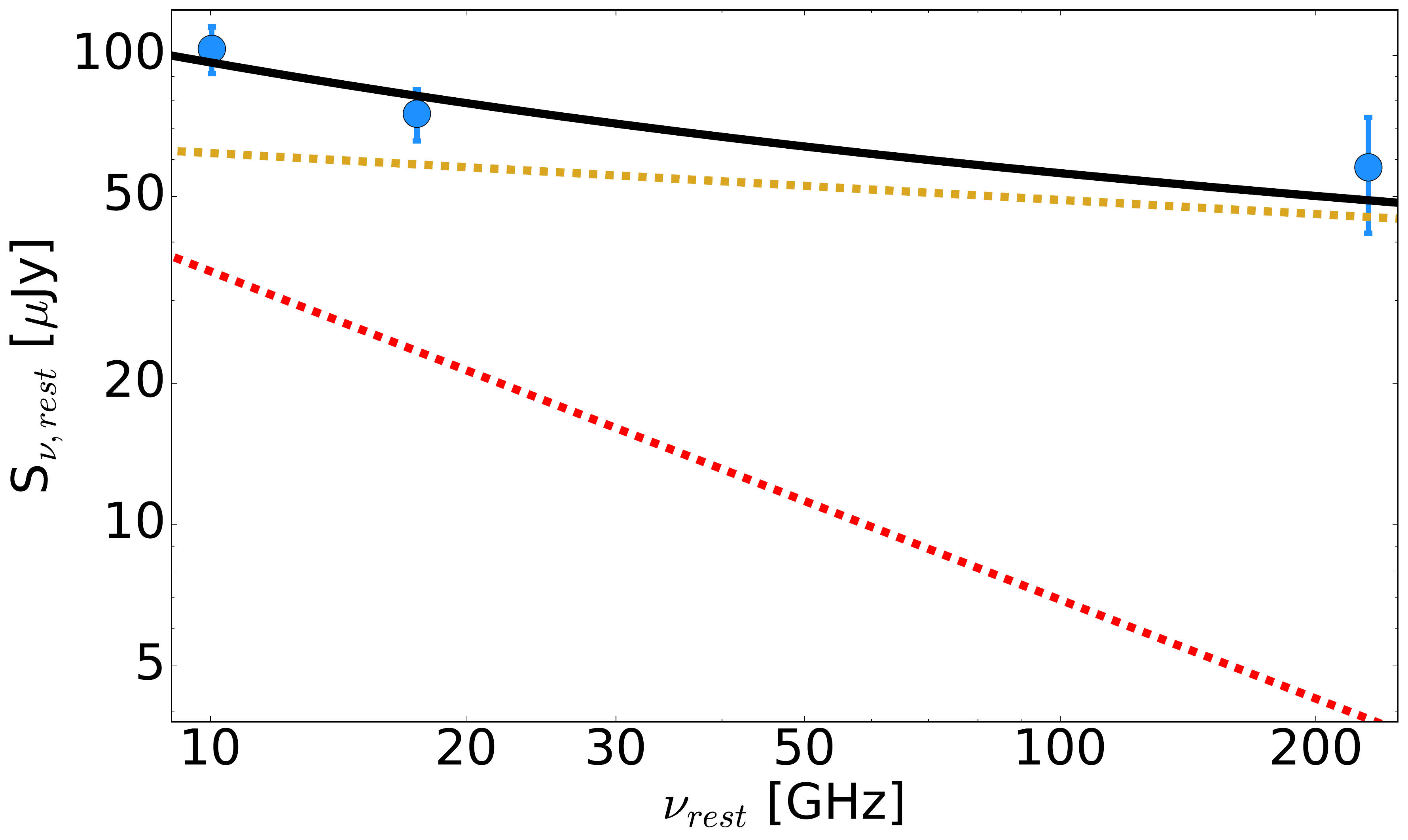}
\caption{Radio SED of the core component of 3C 368, decomposed into thermal (gold) and non-thermal (red) contributions. Data from \cite{Best1998} and \cite{Lamarche2017}.}
\end{figure}

\subsubsection{Metallicity From Free-Free Continuum and Far-IR Fine-Structure Lines}
Having determined that the radio SED of the star-forming core of 3C 368 is dominated by thermal free-free continuum at 230 GHz (rest-frame), we can estimate the gas-phase oxygen abundance in 3C 368 using our previous observations of the [O\,{\sc iii}] 52\,$\micron$ line \citep{Lamarche2017}.

The method for determining gas-phase absolute ionic abundances using fine-structure line emission and radio free-free continuum has been applied historically to galactic H\,{\sc ii} regions \citep[e.g.,][]{Herter1981, Rudolph1997} and more recently to high-redshift galaxies \citep[e.g.,][]{Lamarche2018}. This technique provides an alternative to optically-derived metallicities, which suffer both from uncertainties in the temperature structure of H\,{\sc ii} regions, as well as dust extinction of the line flux. Dust obscuration is important in the local Universe, especially in the more luminous star-forming galaxies, and becomes an increasingly important effect in the epoch of peak cosmic star formation, at redshift $\sim$ 1 -- 3. It can be shown that the abundance of ion X$^{i+}$, relative to hydrogen, is given by:

\begin{equation}
\frac{n_{X^{i+}}}{n_{H^+}} = \frac{F_\lambda}{S_{\nu,r}} \frac{4.21 \times 10^{-16} T_{4}^{-0.35} \nu_{9}^{-0.1}}{\epsilon_\lambda} \left( \frac{n_e}{n_p} \right)
\end{equation}

where $F_\lambda$ is the fine-structure line flux in units of erg s$^{-1}$ cm$^{-2}$, S$_{\nu,r}$ is the rest-frame radio free-free flux density at rest-frequency $\nu$ in units of Jy, T$_4$ is the electron temperature in units of 10$^4$ K, $\nu_9$ is the radio emission rest-frequency in units of GHz, n$_e$/n$_p$ is the electron to proton number density ratio, which accounts for the contribution of electrons from non-hydrogen atoms present in the H\,{\sc ii} regions, and $\epsilon_\lambda$ is the emissivity per unit volume of the fine-structure line at wavelength $\lambda$, given by: 

\begin{equation}
\epsilon_\lambda = \frac{j_{ul}}{n_i n_e} = \frac{h \nu_{ul} A_{ul} n_u}{n_e n_i},
\end{equation}

where h is the Planck constant, A$_{ul}$ is the Einstein A coefficient for the line transition at frequency $\nu_{ul}$, 9.8 $\times$ 10$^{-5}$ s$^{-1}$ and 5785.9 GHz, respectively, for the [O\,{\sc iii}] 52\,$\micron$ line \citep{Carilli2013}, and the ratio of the level population in the upper state (n$_u$) to the total (n$_i$) is determined by detailed balance. Adopting the collisional coefficients of \cite{Palay2012}, and an H\,{\sc ii} region electron density of 1,000 cm$^{-3}$, as has been determined to be the case in 3C 368 \citep{Lamarche2017}, we calculate a value of $\epsilon_\lambda$ =  7.6 $\times$ 10$^{-22}$ erg s$^{-1}$ cm$^{3}$ for the [O\,{\sc iii}] 52\,$\micron$ line. We additionally assume n$_e$/n$_p$ = 1.05, which accounts for the electrons contributed from helium, and T$_4$ = 1, a typical value for H\,{\sc ii} regions.

Combining the total integrated [O\,{\sc iii}] 52\,$\micron$ line flux \citep[1.34 x 10$^{-17}$ W m$^{-2}$;][]{Lamarche2017} with the estimated free-free flux density at 230 GHz (rest-frame), and using Equation 2, we calculate [O$^{++}$/H] = 1 $\times$ 10$^{-4}$. Then, to determine the [O$^{++}$/O] ratio, and hence scale our ionic oxygen abundance to total oxygen abundance, an estimate for the hardness of the ambient radiation field within 3C 368 is required. Using the H\,{\sc ii} region models of \cite{Rubin1985}, we predicted that $\sim$ 84\% of the oxygen within the H\,{\sc ii} regions of 3C 368 is in the O$^{++}$ state \citep{Lamarche2017}. Using that ionization fraction here, we obtain an estimated [O/H] ratio of 1.2 $\times$ 10$^{-4}$, or $\sim$ 0.3 Z$_\odot$ \citep{Asplund2009}, within about a factor of two uncertainty, where this uncertainty is dominated by both the uncertainty in determining the free-free contribution to the radio SED and the uncertainty in the measured [O\,{\sc iii}] 52\,$\micron$ line flux.

This low-metallicity value is consistent with the results of \cite{Croxall2017}, who found empirically that the [C\,{\sc ii}] 158\,$\micron$ / [N\,{\sc ii}] 205\,$\micron$ line ratio is highly elevated in low-metallicity sources. In \cite{Lamarche2017} we found a lower-bound of [C\,{\sc ii}] 158\,$\micron$ / [N\,{\sc ii}] 205\,$\micron$  \textgreater\,60, which would indicate 12 + log(O/H) $\sim$ 8.1, or equivalently Z $\sim$ 0.25 Z$_\odot$, in the sample from \cite{Croxall2017}.

An alternative explanation for the radio emission emanating from the core component of 3C 368 is that it originates from the AGN itself. Strong synchrotron self-absorption can lead to AGN with flat radio spectra, such that the radio emission which we attribute to thermal free-free emission from extended star-forming H\,{\sc ii} regions may instead come from a compact AGN \citep[e.g.,][]{Roellig1986}. As a check on this possibility, we examine the radio/IR correlation, q$_{IR}$ \citep[e.g.,][]{Ivison2010}, where

\begin{equation}
q_{IR} \equiv log_{10} \Bigg( \frac{S_{IR}/(3.75\,\times\,10^{12}\,W\,m^{-2})}{S_{1.4\,GHz}/(W\,m^{-2}\,Hz^{-1})} \Bigg),
\end{equation}

S$_{IR}$ is the total integrated IR flux (rest-frame 8 -- 1000\,$\micron$) in units of W\,m$^{-2}$, and S$_{1.4\,GHz}$ is the flux density at 1.4 GHz in units of W\,m$^{-2}$\,Hz$^{-1}$. Extrapolating the SED-decomposed thermal contribution to the radio emission at rest-frame 230\,GHz (45\,$\mu$Jy) to the expected value at 1.4 GHz using a free-free power-law index of -0.1, and using the total FIR luminosity attributed to star formation in 3C 368 \citep[2.0 $\times$ 10$^{12}\,L_{\odot}$,][]{Podigachoski2015} to calculate S$_{IR}$, we obtain a value of q$_{IR}$ = 2.6. This value is within the range seen in star-forming galaxies \citep[e.g.,][]{Ivison2010}, suggesting that the attribution of the radio continuum to free-free emission in 3C 368 is appropriate.

Unfortunately, our 230\,GHz rest-frame ALMA observations \citep{Lamarche2017}, with a synthesized beam of size 3$\farcs$12 $\times$ 1$\farcs$81, do not resolve the core of 3C 368. We will propose for higher resolution continuum imaging with ALMA to test the theory of extended H\,{\sc ii} regions vs. compact AGN as the source for the high-frequency radio continuum in 3C 368, thereby supporting or disfavoring the low-metallicity hypothesis for the ISM within 3C 368.

\subsubsection{Comparison With Optically-Derived Metallicity}
The sub-solar oxygen abundance that we derive for 3C 368 using the [O\,{\sc iii}] 52\,$\micron$ line and radio free-free continuum can be compared to the same value derived using optical lines. Observations exist for 3C 368 in the [O\,{\sc iii}] 5007\,\AA, [O\,{\sc ii}] 3727\,\AA, and H$\delta$ lines, such that R$_{23}$ \citep[e.g.,][]{Pagel1979} can be calculated by assuming a scaling from H$\delta$ to H$\beta$, where

\begin{equation}
R_{23} \equiv \frac{[O II] \lambda 3727 + [O III] \lambda \lambda 4959,5007}{H\beta}.
\end{equation}

Adopting an [O\,{\sc iii}] 5007\,\AA\, line flux of 6.8 $\times$ 10$^{-18}$ W m$^{-2}$ \citep{Jackson1997}, an [O\,{\sc ii}] 3727\,\AA\, line flux of 5.9 $\times$ 10$^{-18}$ W m$^{-2}$ \citep{Best2000}, an H$\delta$ line flux of 2.8 $\times$ 10$^{-19}$ W m$^{-2}$ \citep{Best2000}, and assuming the theoretical scaling of H$\delta$/H$\beta$ = 0.26 \citep[Case-B recombination, e.g.,][]{AGN2}, we obtain an R$_{23}$ value of $\sim$ 12. This R$_{23}$ value indicates 12 + log(O/H) $\sim$ 8.3 \citep{Kewley2002}, or $\sim$ 0.4 Z$_\odot$ \citep{Asplund2009}, making it consistent with the FIR/radio-derived estimate.

\subsection{PDRs}

\subsubsection{PDR Parameters}

In \cite{Lamarche2017}, we modeled the photo-dissociation regions (PDRs) within 3C 368 utilizing the PDR Toolbox \citep{Pound2008, Kaufman2006}, together with detections of the [C\,{\sc ii}] 158 $\micron$ \citep{Stacey2010CII} and [O\,{\sc i}] 63 $\micron$ lines, as well as the modeled far-IR continuum luminosity \citep{Podigachoski2015}, obtaining best-fit parameters of G$_0$ $\sim$ 280 Habing units and n $\sim$ 7,500 cm$^{-3}$. Here we update the model to include the two newly-detected lines.

Including the CO(4-3) and [C\,{\sc i}] 609 $\micron$ lines, we obtain new best-fit values of G$_0$ $\sim$ 320 Habing units and n $\sim$ 3,200 cm$^{-3}$, with a PDR surface temperature of $\sim$ 200 K (see Figure 5). These PDR parameters are similar to those which we obtained previously, up to a decrease in the density, due to the detection significance of the CO(4-3) and [C\,{\sc i}] 609 $\micron$ lines, which tightly constrain that parameter. We again see that the [C\,{\sc ii}] emission is elevated relative to the CO emission, and now also relative to the [C\,{\sc i}] emission. 

The penetration of far-UV photons capable of photodissociating CO and ionizing carbon is limited by dust extinction. Therefore, assuming dust content scales with metallicity, the CO-photodissociated surface of a molecular cloud becomes much larger relative to the shielded CO-emitting core in low-metallicity environments. In PDR models \citep[e.g.,][]{Kaufman2006}, [C\,{\sc i}] line emission arises from a thin transition region of neutral carbon that lies between the C$^+$ and CO regions. The [C\,{\sc ii}] line flux is therefore enhanced relative to both CO and [C\,{\sc i}] in these environments \citep[e.g.,][]{Maloney1988, Stacey1991}.

\begin{figure}
\plotone{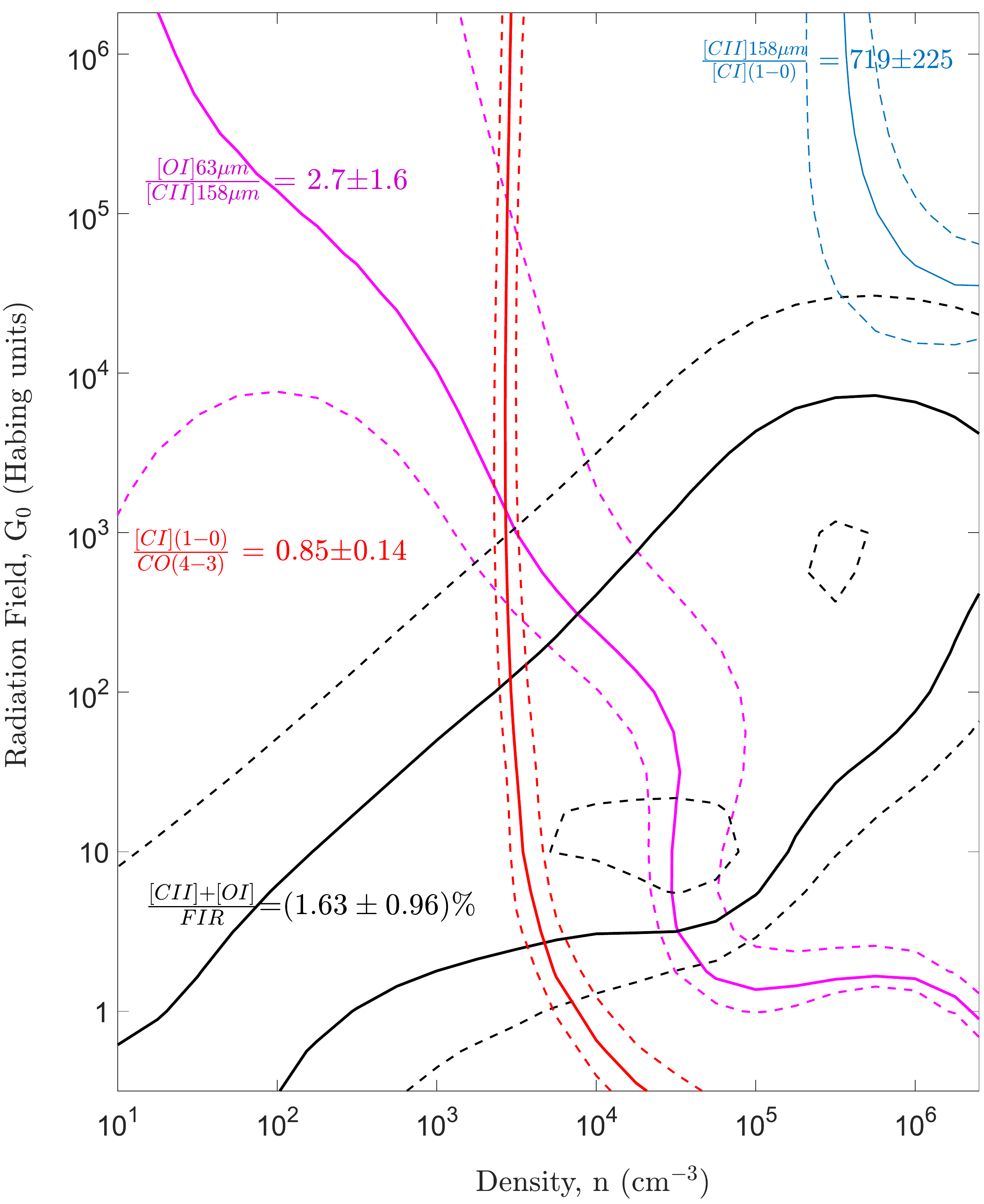}
\caption{A plot of the best-fit PDR parameters in 3C 368, generated using the models of \cite{Pound2008} and \cite{Kaufman2006}. The fit favors PDRs with G$_0$ $\sim$ 320 Habing units and n $\sim$ 3,200 cm$^{-3}$.}
\end{figure}

\subsubsection{PDR Mass}

Using the calculated n, G$_0$, and PDR surface temperature values in 3C 368, we can estimate the PDR mass following \cite{Hailey-Dunsheath2010}:

\begin{equation}
\begin{split}
\frac{M_{PDR}}{M_\odot} = 0.77 \bigg( \frac{0.93 L_{[C\,\textrm{\sc ii}]}}{L_\odot} \bigg) \bigg( \frac{1.4 \times 10^{-4}}{X_{C^+}} \bigg) \\
\times \frac{1 + 2 \exp(\frac{-91 K}{T}) + \frac{n_{crit}}{n}}{2 \exp(\frac{-91 K}{T})},
\end{split}
\end{equation}

where $X_{C^+}$ is the abundance of C$^+$ per hydrogen atom, taken here to be $1.4 \times 10^{-4}$ \citep{Savage1996}, n$_{crit}$ is the critical density of the [C\,{\sc ii}] 158 $\micron$ transition \citep[2,800 cm$^{-3}$,][]{Stacey2011}, and assuming that $\sim$ 93\% of the [C\,{\sc ii}] emission originates within the PDRs of 3C 368 as we estimated in \cite{Lamarche2017}. We calculate a PDR gas mass of $\sim$ 1.7 $\times$ 10$^{10}$ $M_{\odot}$ in this source (see Table 2).

\subsubsection{Molecular-Gas Mass}

With detections of both the CO(4-3) and [C\,{\sc i}] 609 $\micron$ lines, the molecular-gas mass within 3C 368 can be estimated in three ways:

First, using the detection of the CO(4-3) line, we can estimate the molecular-gas mass by assuming a conversion factor back to CO(1-0) and then adopting an $\alpha_{CO}$ value. The measured CO(4-3) line flux of 1.08 $\pm$ 0.14 Jy km s$^{-1}$ equates to a value of L'$_{CO(4-3)}$ = 4.6 $\pm$ 0.6 $\times$ 10$^{9}$ K km s$^{-1}$ pc$^2$. Our previous non-detection of the CO(2-1) line with ALMA yields a 3$\sigma$ upper limit of L'$_{CO(2-1)}$ \textless  3.45 $\times$ 10$^{9}$ K km s$^{-1}$ pc$^2$ \citep{Lamarche2017}. This implies that r$_{43/21}$ \textgreater 1.3, so we use this excitation ratio to scale L'$_{CO(4-3)}$ back to L'$_{CO(1-0)}$. Taking an ultraluminous infrared galaxy (ULIRG) value of $\alpha_{CO}$ = 0.8 $M_{\odot}$(K km s$^{-1}$ pc$^2$)$^{-1}$ \citep[e.g.,][]{Bolatto2013}, we obtain a molecular-gas mass of 2.9 $\times$ 10$^{9}$ $M_{\odot}$.

If, instead, we calculate a value of $\alpha_{CO}$ based on the metallicity estimates from section 3.2, adopting the scaling of \cite{Glover2011} \citep[and see also][]{Bolatto2013}, we obtain $\alpha_{CO}$ $\sim$ 18 $M_{\odot}$(K km s$^{-1}$ pc$^2$)$^{-1}$. Adopting this larger value of $\alpha_{CO}$, and assuming the same super-thermal CO-excitation as above, we obtain a molecular mass of 6.4 $\times$ 10$^{10}$ $M_{\odot}$ (see Table 2). This metallicity-adjusted $\alpha_{CO}$-based molecular gas mass is $\sim$ 4x larger than the PDR gas mass estimated above, making the mass ratio consistent with that observed in other starburst galaxies \citep[e.g.,][]{Stacey1991}, lending further credibility to the low-metallicity explanation for the small CO luminosity in 3C 368.

The second method is based on a scaling of molecular gas mass with [C\,{\sc i}]-line luminosity -- which does not assume that the [C\,{\sc i}] line arises from a thin transition in the PDRs -- as indicated by the work of \cite{Papadopoulos2004Q10}. Following \cite{Bothwell2017}:

\begin{equation}
\begin{split}
M(H_2)^{[CI]} = 1375.8 D_{L}^{2} (1+z)^{-1} \left( \frac{X_{[CI]}}{10^{-5}} \right) ^{-1} \\
\times \left( \frac{A_{10}}{10^{-7} s^{-1}} \right) ^{-1} Q_{10}^{-1} S_{[CI]} \Delta v ,
\end{split}
\end{equation}

where D$_L$ is the luminosity distance in Mpc (7,735 in the case of 3C 368), X$_{[CI]}$ is the [C\,{\sc i}]/H$_2$ abundance ratio, taken here to be 3 $\times$ 10$^{-5}$ \citep[e.g.,][]{Weiss2003, Papadopoulos2004}, A$_{10}$ is the Einstein A coefficient, 7.93 $\times$ 10$^{-8}$ s$^{-1}$ \citep{Wiese1966}, Q$_{10}$ is the excitation parameter which depends on gas density, kinetic temperature and radiation field \citep[e.g.,][]{Papadopoulos2004Q10}, which we take to be 0.6 \citep{Bothwell2017}, and S$_{[CI]}$ $\Delta$v is the flux of the [C\,{\sc i}] 609 $\micron$ line in units of Jy km s$^{-1}$. We obtain a molecular-gas mass of 2.3 $\times$ 10$^{10}$ $M_{\odot}$. We note that this molecular-gas mass is $\sim$ 3x smaller than that derived using the metallicity-corrected $\alpha_{CO}$ method above. If we scale the [C\,{\sc i}]/H$_2$ abundance ratio, X$_{[CI]}$, by the sub-solar gas-phase metallicity estimated in section 3.2, Z $\sim$ 0.3 Z$_\odot$, we calculate an H$_2$ mass of 7.7 $\times$ 10$^{10}$ $M_{\odot}$. Boosting this mass by $\sim$ 20\% to account for the contribution from helium, which is already accounted for in the $\alpha_{CO}$-based method above, increases the mass estimate to 9.2 $\times$ 10$^{10}$ $M_{\odot}$, such that the two methods produce masses which differ by $\sim$ 30\%.

Finally, we can use the dust mass calculated by multi-component SED modeling, along with a gas-to-dust ratio, $\delta_{GDR}$, to estimate the molecular-gas mass within 3C 368. Adopting an SED-modeled dust-mass of 1.4 $\times$ 10$^{8}$ $M_{\odot}$ \citep{Podigachoski2015} and a gas-to-dust ratio of 540, appropriate for Z $\sim$ 0.3 Z$_\odot$ \citep{Remy-Ruyer2014}, we obtain a molecular-gas mass of 7.6 $\times$ 10$^{10}$ $M_{\odot}$. Or, equivalently, if we use the CO-derived gas-mass together with the SED-modeled dust-mass, we obtain a value of $\delta_{GDR}$ = 464 (see Table 2). Since $\alpha_{CO}$ and $\delta_{GDR}$ scale differently with metallicity, the CO- and dust-derived gas masses are incompatible at solar metallicity, giving us more confidence that 3C 368 does indeed have sub-solar metallicity, and that our derived gas masses are robust.

The calculated molecular-gas mass implies that, at its current star-formation rate, $\sim$ 350 $M_{\odot}$ yr$^{-1}$ \citep{Podigachoski2015}, 3C 368 will deplete its supply of molecular gas in $\sim$ 170 Myrs. We can additionally calculate the gas fraction, f$_{gas} \equiv \frac{M_{gas}}{M_{gas} + M_{stars}}$, and the specific star-formation rate, sSFR $\equiv \frac{SFR}{M_{stars}}$, in 3C 368, finding values of $\sim$ 0.15 and $\sim$ 1 Gyr$^{-1}$, respectively (see Table 2). The calculated sSFR puts 3C 368 on the upper end of the galaxy main-sequence as defined by \cite{Genzel2015}, with SFR/SFR$_{MS}$ = 2.8, and the gas fraction lies exactly on the scaling with sSFR/sSFR$_{MS}$, stellar mass, and redshift found in \cite{Scoville2017}. Moreover, 3C 368 lies along the L'$_{[CI]}$ to L$_{IR}$ correlation found in other high-redshift main-sequence galaxies, however it is elevated in its L'$_{[CI]} / $L'$_{CO(2-1)}$ ratio, with a lower limit of 0.9, by a factor of $\sim$ 4.4 \citep{Valentino2018}. The models of \cite{Papadopoulos2018} find the [C\,{\sc i}](1-0)/CO(1-0) brightness ratio to be much larger than unity in low-metallicity or cosmic-ray-dominated regions, however, since we have only a modest lower limit on this ratio, we cannot draw any strong conclusions from it. When compared to a sample of low-metallicity high-redshift sources \citep{Coogan2019}, 3C 368 is observed to be similarly deficient in CO emission, and when compared to a sample of local luminous infrared galaxies (LIRGs) and spirals \citep{Liu2015}, it has a CO(4-3) flux which falls below that expected from the scaling with FIR luminosity, by a factor of $\sim$ 3.7.

\subsection{Dynamical Mass}

Since our ALMA observations of 3C 368 spatially resolve the source, we can additionally estimate the dynamical mass of the galaxy. Assuming a disk geometry and circular orbits:

\begin{equation}
M_{dyn} = \frac{v^2_{rot} r}{G},
\end{equation}

where v$_{rot}$ is the true rotational velocity of the disk, estimated from the observed velocity by correcting for the average inclination angle, $\langle$v$_{rot}$$\rangle$ $\sim$ $\frac{\pi}{2}$ v$_{obs}$ \citep[e.g.,][]{Erb2006}, r is the radius of the line-emitting region of the galaxy, and G is the universal gravitational constant.

We estimate the radius from two-dimensional Gaussian fits to the CO(4-3) moment-zero map (r = 0.5\,FWHM = 3.14 kpc). Similarly, we take the observed velocity to be half the full-width-half-maximum (FWHM) of the CO(4-3) line (v$_{obs}$ = 284 km s$^{-1}$). We obtain a dynamical mass of $\sim$ 1.5 $\times$ 10$^{11}$ $M_{\odot}$ (see Table 2). This dynamical mass estimate is $\sim$ 2x smaller than the estimated stellar mass within 3C 368 \citep[$\sim$ 3.6 $\times$ 10$^{11}$ M$_\odot$,][]{Best1998stellarmass}, a discrepancy that can be due to the effects of inclination angle and mass profile on the calculated dynamical mass.

\begin{deluxetable}{ccc}
\tablecaption{Physical Properties of 3C 368}
\tablecolumns{3}
\tablenum{2}
\tablehead{
\colhead{Property} & \colhead{Value} & \colhead{Reference}
}
\startdata
R.A. (J2000) & 18$^h$05$^m$06$^s$.40 & \cite{Best1998} \\
Dec. (J2000) & +11$^o$01'33".09 & \cite{Best1998} \\
z & 1.131 & \cite{Meisenheimer1992} \\
SFR & 350 $M_{\odot}$ yr$^{-1}$ & \cite{Podigachoski2015} \\
M$_*$ & 3.6 $\times$ 10$^{11}$ M$_\odot$ & \cite{Best1998stellarmass} \\
M$_{gas}$ & 6.4 $\times$ 10$^{10}$ $M_{\odot}$ & This Work \\
M$_{PDR}$ & 1.7 $\times$ 10$^{10}$ $M_{\odot}$ & This Work \\
M$_{dust}$ & 1.4 $\times$ 10$^{8}$ $M_{\odot}$ & \cite{Podigachoski2015} \\
M$_{dyn}$ & \textgreater 1.5 $\times$ 10$^{11}$ $M_{\odot}$ & This Work \\
$\delta_{GDR}$ & 464 & This Work \\
f$_{gas}$ & 0.15 & This Work \\
sSFR & 1 Gyr$^{-1}$ & This Work \\
SFR/SFR$_{MS}$ & 2.8 & This Work, \cite{Genzel2015} \\
$\tau_{Depletion}$ & 170 Myr & This Work \\
\enddata
\tablecomments{See the text for explanations of the physical properties presented in this work.}
\end{deluxetable}

\section{Conclusions}

We have presented detections of the CO(4-3) and [C\,{\sc i}] 609 $\micron$ spectral lines, as well as the rest-frame 480.5\,GHz dust continuum, in 3C 368, an FR-II type galaxy at redshift 1.131.

We have estimated the gas-phase metallicity in 3C 368 using two independent methods: (1) combining the observed [O\,{\sc iii}] 52\,$\micron$ fine-structure line flux with radio free-free continuum emission and (2) employing the optical [O\,{\sc iii}] 5007\,\AA\, and [O\,{\sc ii}] 3727\,\AA\, lines (R$_{23}$ technique). Both methods arrive at a consistent gas-phase metallicity of $\sim$ 0.3 Z$_\odot$.

We have calculated the molecular-gas mass within 3C 368 in three independent ways: (1) using the CO(4-3) line luminosity, gas excitation, and a metallicity-adjusted $\alpha_{CO}$ conversion factor, (2) scaling from the luminosity of the atomic carbon line, and (3) adopting a conversion factor from dust mass to molecular-gas mass. We find a consistent gas mass estimate of $\sim$ 6.4 $\times$ 10$^{10}$ $M_{\odot}$ across these three methods only when adopting a gas-phase metallicity of $\sim$ 0.3 Z$_\odot$.

Considering this molecular-gas mass, and a star-formation rate of $\sim$ 350 $M_{\odot}$ yr$^{-1}$ \citep{Podigachoski2015}, we deduce that 3C 368 has a molecular-gas depletion-time, $\tau_{Depletion}$, of $\sim$ 170 Myrs. We also calculate a gas fraction, f$_{gas}$, and specific star-formation rate, sSFR, adopting a stellar mass of $\sim$ 3.6 $\times$ 10$^{11}$ M$_\odot$ \citep{Best1998stellarmass}, of $\sim$ 0.15 and $\sim$ 1 Gyr$^{-1}$, respectively. The calculated sSFR puts 3C 368 on the upper end of the galaxy main-sequence, as defined by \cite{Genzel2015}, and the gas fraction lies exactly on the scaling with sSFR/sSFR$_{MS}$, stellar mass, and redshift found in \cite{Scoville2017}.

Further observations will allow us to determine whether 3C 368 does indeed have low gas-phase metallicity, if indeed free-free continuum emission dominates the radio spectrum at high frequencies, or whether the AGN is strongly contaminating the emission in those bands. In particular, we will propose for higher angular-resolution observations at rest-frame 230\,GHz to determine whether the radio continuum is extended, as should be the case if it originates from star-forming H\,{\sc ii} regions, or compact, as would be the case if it is of AGN origin.

\section*{Acknowledgments}

We thank the anonymous referee for the insightful comments and detailed suggestions that helped to improve this manuscript. C.L. acknowledges support from an NRAO Student Support Award, SOSPA3-011, and from NASA grant NNX17AF37G. D.B. acknowledges support from FONDECYT postdoctorado project 3170974.

This paper makes use of the following ALMA data: ADS/JAO.ALMA$\#$2015.1.00914.S. ALMA is a partnership of ESO (representing its member states), NSF (USA) and NINS (Japan), together with NRC (Canada) and NSC and ASIAA (Taiwan) and KASI (Republic of Korea), in cooperation with the Republic of Chile. The Joint ALMA Observatory is operated by ESO, AUI/NRAO and NAOJ.

The National Radio Astronomy Observatory is a facility of the National Science Foundation operated under cooperative agreement by Associated Universities, Inc.

\bibliography{3C368_updated_bib_mod_mod.bib}

\end{document}